\documentclass[conference]{IEEEtran}
\IEEEoverridecommandlockouts

\usepackage{cite}
\usepackage{amsmath,amssymb,amsfonts}
\usepackage{graphicx}
\usepackage{textcomp}
\usepackage{xcolor}
\usepackage{booktabs}
\usepackage{array}
\usepackage{multirow}
\usepackage{url}

\begin{document}

\title{How Robust Are Neural Audio Codecs for African Speech?\\ A Multi-Task Benchmark and the Limits of Perceptual Quality}


\author{
  \IEEEauthorblockN{Chibuzor Okocha and Christian Grant}
  \IEEEauthorblockA{\textit{Department of Computer \& Information Science \& Engineering} \\
  \textit{University of Florida}, Gainesville, FL, USA \\
  \{c.okocha, christan\}@ufl.edu}
}

\maketitle

\begin{abstract}
Neural audio codecs enable low-bitrate speech compression and tokenization, yet their robustness on accented and multilingual speech remains under-evaluated. We benchmark seven open-source neural codecs (DAC, EnCodec, FocalCodec, LanguageCodec, SemantiCodec, UniCodec, WavTokenizer) on three African speech datasets (\textit{afrinames}, \textit{afrispeech\_dialog}, \textit{afrispeech\_multilingual}), reporting signal-level quality (NISQA, UTMOS, ViSQOL, STOI, F0-RMSE) alongside two downstream tasks: automatic speech recognition (ASR) and speaker verification (ASV). Our analysis yields four findings. First, signal metrics differ sharply in downstream validity: reference-based structural/intelligibility measures (ViSQOL, STOI) and prosodic error (F0-RMSE) track ASR and ASV degradation far more reliably than neural mean-opinion-score predictors (NISQA, UTMOS). Second, intelligibility and speaker-identity preservation diverge sharply across architectures, and the apparent identity ranking itself depends on the ASV backend. Third, degradation is strongly domain-dependent and largest for conversational dialog. Fourth, the resulting degradation is partly recoverable: parameter-efficient \emph{codec} adaptation (LoRA, $\sim$1--3\% of parameters) on roughly 35 hours of African speech reduces the compression-induced word-error-rate gap, recovering recognition toward the uncompressed baseline (developed in companion work). These results motivate task-aware, domain-representative, and adaptation-aware evaluation of speech codecs as a prerequisite for inclusive deployment.
\end{abstract}


\section{Introduction}
The emergence of neural audio codecs has significantly advanced speech compression and generative modeling \cite{defossez2022, kumar2023, okocha_reasoning, ref_afridialog, ref_afrisem, ref_afrivox}. By combining high-capacity neural networks with vector quantization, these models achieve high-fidelity reconstruction at bitrates far below traditional psychoacoustic methods \cite{shi2024a, sadok2025}. However, their development and benchmarking have centered on high-resource languages \cite{wu_codec-superb_2024}, largely neglecting the linguistic and acoustic diversity of the African continent \cite{sanni_afrispeech-dialog_2025}. As the field moves toward more inclusive technology, it is essential to understand how codecs handle the phonetic structures, tonal variation, and spontaneous dialog characteristic of African speech.

We present a systematic benchmark of seven state-of-the-art neural audio codecs---Descript Audio Codec (DAC), EnCodec, FocalCodec, LanguageCodec, SemantiCodec, UniCodec, and WavTokenizer---in the African context. Crucially, we go beyond signal-level quality and ask which evaluation signals actually predict real-world utility. We pair objective signal metrics (NISQA, UTMOS, ViSQOL, STOI, F0-RMSE) \cite{mittag2021, saeki2022, hines2015, taal2011} with two downstream tasks---ASR and speaker verification (ASV)---and we quantify, per dataset, how strongly each signal metric tracks downstream behavior. Finally, because pretrained codecs are not the only deployable artifact, we study whether lightweight adaptation can recover the utility lost to compression.

A consistent picture emerges: the perceptual scores most often used to rank codecs are not the ones that predict downstream success, the relationship is task- and domain-specific, and the degradation introduced by compression is partly recoverable through adaptation. The primary contributions of this work are:
\begin{itemize}
    \item \textbf{A multi-task, multi-domain codec benchmark on African speech.} We evaluate seven codecs on three datasets using signal-level metrics, ASR (Whisper-large-v3 on conversational speech), and ASV across all three domains (Tables~\ref{tab:signal_summary}--\ref{tab:dialog_main}).
    \item \textbf{A metric-validity analysis.} Using per-dataset Spearman correlations (Table~\ref{tab:corr}), we show that reference-based intelligibility/structure metrics and F0 error track downstream degradation far better than the neural MOS predictors (NISQA, UTMOS), and that the most informative metric differs by task and domain.
    \item \textbf{An adaptation outlook.} We summarize a companion result (Table~\ref{tab:ft}) showing that parameter-efficient \emph{codec} adaptation recovers part of the compression-induced ASR gap on conversational speech, reframing codec robustness as partly an adaptation problem rather than a fixed property.
\end{itemize}

\section{Related Work}
Neural audio coding has shifted from psychoacoustic methods toward data-driven architectures that use vector quantization for high-fidelity compression at low bitrates \cite{defossez2022, kumar2023}. EnCodec \cite{defossez2022} and DAC \cite{kumar2023} use residual vector quantization (RVQ) for strong rate--distortion performance. More recent models such as WavTokenizer and SemantiCodec target ultra-low bitrates by extracting semantic tokens from pretrained speech models \cite{ji2025b, liu2024}, while FocalCodec and LanguageCodec emphasize linguistic-feature preservation relevant to low-resource speech \cite{dellalibera2025, ji2025a}.

Although early evaluations relied on signal-level quality, there is growing consensus that objective scores alone do not capture practical utility. Neural quality estimators such as NISQA and UTMOS \cite{mittag2021, saeki2022} are now reported alongside ViSQOL and STOI \cite{hines2015, taal2011} as proxies for human perception, yet how codec artifacts propagate into machine-learning tasks remains under-studied. Our work extends application-aware benchmarking \cite{wu_codec-superb_2024} by measuring, rather than assuming, the link between signal metrics and downstream tasks. African languages remain under-represented in codec training and validation; the tonal and acoustic characteristics in corpora such as \textit{afrispeech\_multilingual} and \textit{afrispeech\_dialog} pose challenges typically absent from high-resource benchmarks. By coupling per-dataset signal quality to ASR and ASV outcomes, we establish a performance baseline for the African setting.

\section{Methodology and Experimental Design}

\subsection{Problem Formulation}
Given a speech signal $x \in \mathbb{R}^T$, a neural audio codec $f_\theta$ encodes $x$ into discrete tokens $z$ and reconstructs $\hat{x} = f_\theta^{-1}(z)$. Most codecs are trained on predominantly Western English corpora. We study two regimes: (i) \emph{zero-shot}, using publicly released pretrained checkpoints without modification, which reflects generalization to African speech as deployed today; and (ii) \emph{adaptation}, in which a lightweight component is fine-tuned on African speech. We ask how much utility is lost in regime (i), and how much of that loss is recovered in regime (ii).

\subsection{Datasets}
Three datasets capture complementary sources of accent and language variation (Table~\ref{tab:dataset_protocol}; Table~\ref{tab:codec_representations} lists codec configurations):
\begin{itemize}
    \item \textbf{\textit{afrinames}} \cite{olatunji2023}: African proper names and short identity-bearing phrases, stressing phonetic and speaker-identity preservation.
    \item \textbf{\textit{afrispeech\_dialog}} \cite{sanni_afrispeech-dialog_2025}: conversational African-accented English with spontaneous dialogue and prosodic variability.
    \item \textbf{\textit{afrispeech\_multilingual}} \cite{ardila2020}: a Common Voice-derived multilingual subset for Nigerian languages, used here primarily for signal and ASV evaluation.
\end{itemize}
All audio was resampled to each codec's required input rate (16 or 24\,kHz). No enhancement or denoising was applied.

\begin{table}[t]
\centering
\scriptsize
\setlength{\tabcolsep}{3pt}
\caption{Dataset and evaluation protocol summary. ``Eval.'' counts are the files scored in the primary downstream runs; ASV counts are exhaustive target/impostor trials.}
\label{tab:dataset_protocol}
\begin{tabular}{@{}lcccl@{}}
\toprule
\textbf{Dataset} & \textbf{Manifest} & \textbf{Eval.} & \textbf{ASV trials} & \textbf{Coverage} \\
 & \textbf{clips} & \textbf{files} & & \\
\midrule
\textit{afrinames} & 280 & 115--150 & 13.2k--22.5k & 8 accents, 24\,kHz \\
\textit{afrispeech\_dialog} & 210 & 49 & 2.4k & 6 accents, 24\,kHz \\
\textit{afrispeech\_multilingual} & -- & 8--100 & 64--10.0k & Common Voice, 24\,kHz mono \\
\bottomrule
\end{tabular}
\\[2pt]
\footnotesize The multilingual manifest was not available in the local benchmark snapshot; evaluated-file counts are therefore taken from the metric outputs. DAC has only 8 multilingual ASV files, while most other codecs have 100.
\end{table}

\subsection{Neural Audio Codecs}
We evaluate seven open-source codecs spanning diverse paradigms: RVQ models \cite{kumar2023, defossez2022} (DAC, EnCodec); semantic-aware codecs \cite{liu2024, ji2025a} (SemantiCodec, LanguageCodec); and tokenization-based and hybrid models \cite{ji2025b, jiang2025, dellalibera2025} (WavTokenizer, UniCodec, FocalCodec).

\begin{table}[t]
  \caption{Neural audio codec configurations. RVQ models are defined by bitrate; semantic and tokenization models by token rate (Hz).}
  \label{tab:codec_representations}
  \centering
  \begin{footnotesize}
  \begin{tabular}{lll}
    \toprule
    \textbf{Codec} & \textbf{Primary Metric} & \textbf{Configs} \\
    \midrule
    DAC \cite{kumar2023} & Bitrate (kbps) & 8, 16, 24 \\
    EnCodec \cite{defossez2022} & Bitrate (kbps) & 3, 6, 12, 24 \\
    FocalCodec \cite{dellalibera2025} & Bitrate (kbps) & 0.16--0.65 \\
    LanguageCodec \cite{ji2025a} & Bitrate (kbps) & 0.3 (target) \\
    SemantiCodec \cite{liu2024} & Token Rate (Hz) & 25, 50, 100 \\
    UniCodec \cite{jiang2025} & Bitrate (kbps) & 3.4, 6.6 \\
    WavTokenizer \cite{ji2025b} & Token Rate (Hz) & 40, 75 \\
    \bottomrule
  \end{tabular}
  \end{footnotesize}
\end{table}

\subsection{Evaluation Metrics}
\textbf{Signal level.} We report NISQA and UTMOS \cite{mittag2021, saeki2022} for neural MOS prediction, ViSQOL \cite{hines2015} for structural/spectral similarity, STOI \cite{taal2011} for intelligibility, and F0-RMSE for prosodic preservation, the latter being especially relevant to tonal languages.

\noindent\textbf{ASR.} We measure intelligibility of reconstructed speech via Word Error Rate (WER) and Character Error Rate (CER). Whisper-large-v3 \cite{radford2023robust} is our primary fixed recognizer, held constant across codec conditions to isolate the effect of compression; the uncompressed baseline on \textit{afrispeech\_dialog} is 17.3 WER. To test whether our conclusions are recognizer-specific, we additionally report parakeet-tdt-0.6b-v2~\cite{sekoyan2025} in the adaptation summary; a full multi-recognizer sweep appears in companion work~\cite{anon_lora}.

\noindent\textbf{ASV.} Speaker-identity preservation is evaluated with Equal Error Rate (EER) and minimum Detection Cost Function (minDCF) under exhaustive same-/different-speaker trials. We adopt ECAPA-TDNN \cite{desplanques2020} speaker embeddings as the primary backend, as modern neural embeddings are far more discriminative than classical representations. To probe how much the backend itself shapes conclusions, we additionally run an MFCC--GMM backend and compare codec rankings between the two (Sec.~\ref{sec:results}). Trial counts range from $\sim$2.4k (\textit{afrispeech\_dialog}) to $\sim$22.5k (\textit{afrinames}) per condition.

\subsection{Adaptation Setup}
\label{sec:adapt}
To test whether the benchmarked degradation is recoverable, we adapt the \emph{codec} (not the recognizer) with parameter-efficient fine-tuning: the pretrained backbone is frozen and LoRA adapters are inserted into encoder and decoder layers. Adaptation uses a 34.6\,h African-accented corpus (10{,}045 train / 1{,}116 validation clips) with speaker-disjoint partitions; evaluation is on held-out \textit{afrispeech\_dialog} audio that is disjoint from the adaptation corpus, so no evaluation utterance is seen during adaptation. We report DAC, the codec that adapts most reliably. The recognizer is held fixed throughout, so any change in WER reflects the codec rather than the ASR model. The full method, additional codec architectures, and architecture-specific stabilization are developed in companion work~\cite{anon_lora}; here we use adaptation only to establish recoverability.

\subsection{Computing Infrastructure}
Inference and fine-tuning used NVIDIA B200 GPUs with PyTorch~2.x and CUDA~12.8. Data loading used torchaudio and Hugging Face \texttt{datasets}. Large-scale evaluation, including exhaustive impostor trials for ASV, was parallelized across compute nodes.

\section{Results and Discussion}
\label{sec:results}
We organize results around three questions: how does signal quality degrade across domains; which signal metrics actually predict downstream utility; and how much of the lost utility can adaptation recover.

\subsection{Signal Quality Is Domain-Dependent}
Table~\ref{tab:signal_summary} reports signal metrics averaged over codec variants per dataset, against the uncompressed reference. Degradation is clearly domain-dependent: conversational \textit{afrispeech\_dialog} has the lowest mean STOI ($0.71$ vs.\ $0.85$ and $0.81$), consistent with spontaneous speech and prosodic variability being hardest to preserve under compression. The multilingual set has the lowest perceptual baseline (UTMOS $1.84$, NISQA $2.35$), reflecting recording conditions and language coverage rather than compression alone. We also observe that no-reference NISQA can exceed the uncompressed reference (e.g., \textit{afrinames}), a known artifact of no-reference MOS predictors that ``clean up'' audio---further reason not to treat such scores as ground-truth utility.

\begin{table}[t]
\centering
\footnotesize
\caption{Signal-level quality per dataset (mean over codec variants) vs.\ the uncompressed reference. ViSQOL/STOI: higher is better; STOI $\in[0,1]$.}
\label{tab:signal_summary}
\setlength{\tabcolsep}{4pt}
\begin{tabular}{@{}llcccc@{}}
\toprule
\textbf{Dataset} & & \textbf{NISQA} & \textbf{UTMOS} & \textbf{ViSQOL} & \textbf{STOI} \\
\midrule
\multirow{2}{*}{\textit{afrinames}}
 & ref.\      & 2.92 & 2.42 & 5.00 & 1.00 \\
 & codecs     & 3.25 & 2.32 & 4.59 & 0.85 \\
\addlinespace
\multirow{2}{*}{\textit{afrispeech\_dialog}}
 & ref.\      & 3.25 & 2.27 & 5.00 & 1.00 \\
 & codecs     & 3.13 & 2.06 & 4.87 & \textbf{0.71} \\
\addlinespace
\multirow{2}{*}{\textit{afrispeech\_multilingual}}
 & ref.\      & 2.35 & 1.84 & 5.00 & 1.00 \\
 & codecs     & 2.74 & 1.78 & 4.88 & 0.81 \\
\bottomrule
\end{tabular}
\end{table}

\subsection{Which Metrics Predict Downstream Utility?}
Table~\ref{tab:corr} reports per-dataset Spearman correlations between each signal metric and each downstream error metric. The pattern is the central message of this paper: \emph{the no-reference neural MOS predictors are not the most reliable proxies for downstream utility}. UTMOS is weak or non-significant in nearly every downstream comparison (e.g., dialog WER $\rho=-0.26$, multilingual minDCF $\rho=0.24$), and NISQA is at best moderate. By contrast, ViSQOL and STOI strongly track ASR error on dialog and ASV error across domains, while F0-RMSE is especially predictive of ECAPA-TDNN ASV degradation. The practical implication is direct: ranking codecs by headline MOS-style scores can hide damaging failure modes, whereas task-aware reference metrics expose them.

\begin{table}[t]
\centering
\footnotesize
\caption{Signal-vs.-downstream Spearman correlations after recomputing ASV rows with ECAPA-TDNN EER/minDCF. Negative $\rho$ means higher signal quality corresponds to lower downstream error; positive F0-RMSE correlations mean larger pitch error corresponds to larger downstream error.}
\label{tab:corr}
\setlength{\tabcolsep}{2.6pt}
\begin{tabular}{@{}llccccc@{}}
\toprule
\textbf{Dataset} & \textbf{Down.} & \textbf{NISQA} & \textbf{UTMOS} & \textbf{ViSQOL} & \textbf{STOI} & \textbf{F0} \\
\midrule
\textit{dialog} & WER    & $-0.43$ & $-0.26$ & $\mathbf{-0.74}$ & $-0.72$ & $\mathbf{0.85}$ \\
\textit{dialog} & CER    & $-0.43$ & $-0.27$ & $\mathbf{-0.74}$ & $-0.72$ & $\mathbf{0.85}$ \\
\textit{dialog} & EER    & $-0.46$ & $0.15$  & $-0.57$ & $-0.54$ & $\mathbf{0.64}$ \\
\textit{dialog} & minDCF & $-0.56$ & $0.10$  & $-0.65$ & $-0.63$ & $\mathbf{0.72}$ \\
\addlinespace
\textit{afrinames} & EER    & $0.49$ & $0.32$ & $-0.93$ & $-0.89$ & $\mathbf{0.95}$ \\
\textit{afrinames} & minDCF & $0.59$ & $0.45$ & $-0.83$ & $-0.79$ & $\mathbf{0.86}$ \\
\addlinespace
\textit{multiling.} & EER    & $0.37$ & $0.34$ & $\mathbf{-0.93}$ & $-0.93$ & $0.82$ \\
\textit{multiling.} & minDCF & $0.27$ & $0.24$ & $\mathbf{-0.95}$ & $-0.94$ & $0.87$ \\
\bottomrule
\end{tabular}
\\[2pt]
\footnotesize ASR rows use Whisper-large-v3 on \textit{afrispeech\_dialog} ($n=30$; F0 $n=23$). ASV rows use ECAPA-TDNN ($n=26$ for \textit{afrinames}, $n=30$ for dialog/multilingual; F0 $n=19$, 23, and 8 respectively). Bold marks the largest absolute correlation in each row.
\end{table}

\noindent\textbf{Metric masking is visible at the codec level.} The aggregate correlations are not merely statistical artifacts. Table~\ref{tab:masking} shows representative cases where a MOS-style metric ranks a codec near the top, but the same codec is among the worst by downstream utility. WavTokenizer S-600 is ranked first by NISQA on dialog while producing one of the worst WER/CER values; FocalCodec 12.5\,Hz is ranked first by UTMOS on two ASV domains while landing near the bottom by ECAPA EER. These are precisely the failure modes a single headline MOS score would miss.

\begin{table}[t]
\centering
\scriptsize
\setlength{\tabcolsep}{3pt}
\caption{Representative metric-masking cases. Low signal rank is good, but high downstream rank is bad; MMG is the rank gap.}
\label{tab:masking}
\begin{tabular}{@{}llcccr@{}}
\toprule
\textbf{Domain} & \textbf{Codec/variant} & \textbf{Signal} & \textbf{Down.} & \textbf{Ranks} & \textbf{MMG} \\
\midrule
multiling. & Focal 12.5\,Hz & UTMOS & EER & 1 / 31 & 30 \\
afrinames & Focal 12.5\,Hz & UTMOS & EER & 1 / 31 & 30 \\
dialog & WavTok S-600 & NISQA & WER & 1 / 28 & 27 \\
dialog & WavTok S-600 & NISQA & CER & 1 / 28 & 27 \\
\bottomrule
\end{tabular}
\end{table}

\subsection{ASR and ASV Diverge Across Architectures}
\begin{figure}[t]
\centering
\includegraphics[width=\columnwidth]{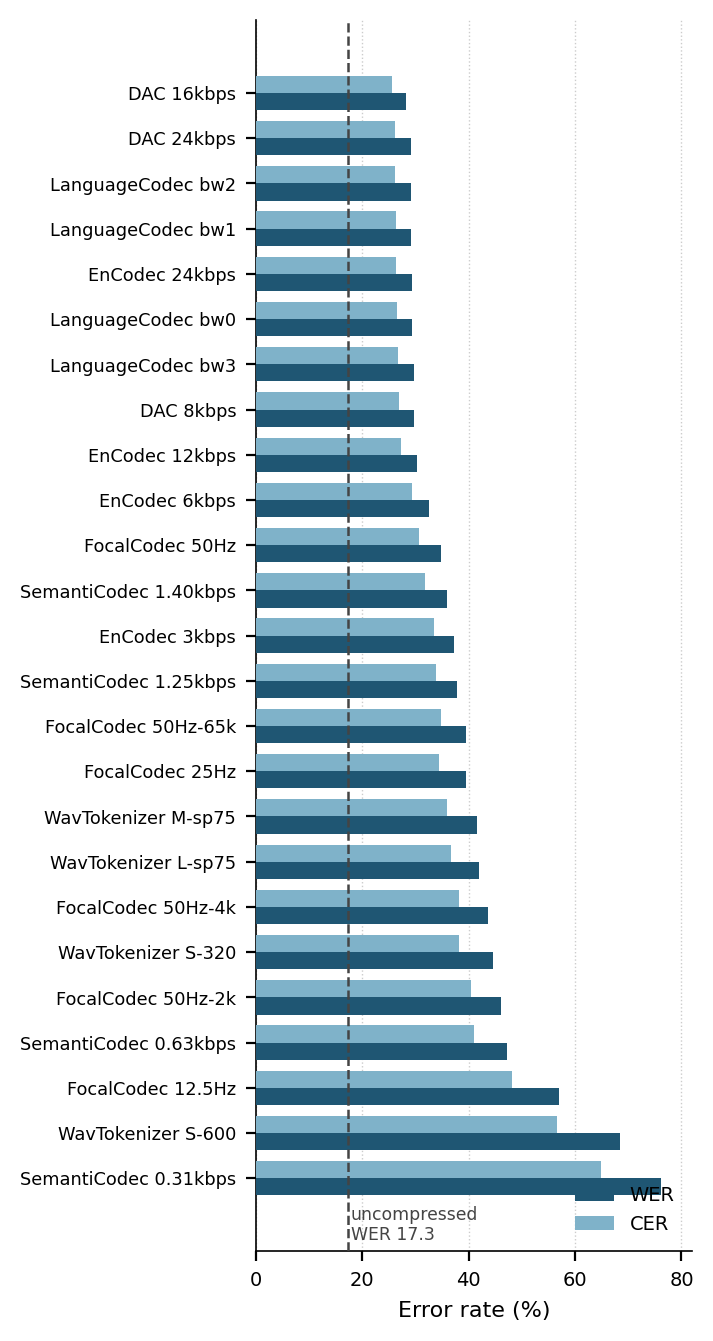}
\caption{Per-variant ASR error on \textit{afrispeech\_dialog} (Whisper-large-v3), sorted by WER. Mid-bitrate RVQ codecs (DAC, EnCodec, LanguageCodec) cluster near the uncompressed baseline (dashed), whereas aggressive semantic and low-token-rate settings degrade severely; CER tracks WER closely.}
\label{fig:dialog_wer}
\end{figure}

Table~\ref{tab:dialog_main} reports per-variant downstream behavior on \textit{afrispeech\_dialog} (Whisper-large-v3 for ASR; ECAPA-TDNN for ASV). All codecs increase WER over the 17.3 uncompressed baseline, but the extent varies widely (Fig.~\ref{fig:dialog_wer}). Mid-to-high-bitrate RVQ settings are most stable (DAC WER 28.3--29.7; LanguageCodec $\approx$29--30), whereas aggressive low-bitrate semantic compression is brittle (SemantiCodec at 0.31\,kbps exceeds 76 WER). Critically, ASR and ASV failure modes diverge: FocalCodec and the low-rate WavTokenizer settings remain usable for recognition (WER 35--57) yet degrade verification badly (EER 4--16), while DAC and LanguageCodec preserve identity perfectly (EER $0$) and UniCodec reverses the pattern---moderate elsewhere but the worst verifier on dialog (EER 10.2). This divergence is exactly what the metric-validity analysis predicts---intelligibility and identity are governed by different signal properties---so codecs cannot be ranked by a single number.

\begin{table}[t]
\centering
\scriptsize
\setlength{\tabcolsep}{3.2pt}
\renewcommand{\arraystretch}{1.05}
\caption{Downstream behavior on \textit{afrispeech\_dialog}: speaker verification (ECAPA-TDNN) vs.\ recognition (Whisper-large-v3; uncompressed baseline 17.3 WER). EER and WER/CER in \%. Lower is better ($\downarrow$). Several ASR-usable codecs still fail verification.}
\label{tab:dialog_main}
\begin{tabular}{@{}lcccc@{}}
\toprule
\textbf{Codec/Var.} & \textbf{EER} $\downarrow$ & \textbf{minDCF} $\downarrow$ & \textbf{WER} $\downarrow$ & \textbf{CER} $\downarrow$ \\
\midrule
\textbf{DAC} \\
\hspace{0.4em}\texttt{16kbps} & 0.00 & 0.00 & 28.26 & 25.54 \\
\hspace{0.4em}\texttt{24kbps} & 0.00 & 0.00 & 29.23 & 26.25 \\
\hspace{0.4em}\texttt{8kbps}  & 0.00 & 0.00 & 29.74 & 26.86 \\
\addlinespace\midrule
\textbf{EnCodec} \\
\hspace{0.4em}\texttt{12kbps} & 0.40 & 0.18 & 30.34 & 27.29 \\
\hspace{0.4em}\texttt{24kbps} & 0.30 & 0.14 & 29.36 & 26.35 \\
\hspace{0.4em}\texttt{3kbps}  & 2.04 & 0.49 & 37.27 & 33.48 \\
\hspace{0.4em}\texttt{6kbps}  & 1.81 & 0.27 & 32.52 & 29.33 \\
\addlinespace\midrule
\textbf{FocalCodec} \\
\hspace{0.4em}\texttt{12.5Hz} & 10.63 & 0.96 & 57.02 & 48.18 \\
\hspace{0.4em}\texttt{25Hz}   & 6.42  & 0.98 & 39.51 & 34.49 \\
\hspace{0.4em}\texttt{50Hz}   & 4.08  & 0.92 & 34.90 & 30.61 \\
\hspace{0.4em}\texttt{50Hz-2k} & 8.16 & 0.88 & 46.10 & 40.56 \\
\hspace{0.4em}\texttt{50Hz-4k} & 6.12 & 0.94 & 43.59 & 38.23 \\
\hspace{0.4em}\texttt{50Hz-65k} & 4.08 & 0.88 & 39.48 & 34.84 \\
\addlinespace\midrule
\textbf{LanguageCodec} \\
\hspace{0.4em}\texttt{bw0} & 0.00 & 0.00 & 29.47 & 26.50 \\
\hspace{0.4em}\texttt{bw1} & 0.00 & 0.00 & 29.29 & 26.30 \\
\hspace{0.4em}\texttt{bw2} & 0.00 & 0.00 & 29.28 & 26.25 \\
\hspace{0.4em}\texttt{bw3} & 0.00 & 0.00 & 29.73 & 26.78 \\
\addlinespace\midrule
\textbf{SemantiCodec} \\
\hspace{0.4em}\texttt{0.31kbps} & 3.61 & 0.74 & 76.22 & 64.96 \\
\hspace{0.4em}\texttt{0.63kbps} & 1.57 & 0.43 & 47.14 & 41.12 \\
\hspace{0.4em}\texttt{1.25kbps} & 0.13 & 0.25 & 37.77 & 33.87 \\
\hspace{0.4em}\texttt{1.40kbps} & 0.11 & 0.19 & 35.91 & 31.91 \\
\addlinespace\midrule
\textbf{WavTokenizer} \\
\hspace{0.4em}\texttt{L-sp75} & 3.68 & 0.70 & 42.04 & 36.63 \\
\hspace{0.4em}\texttt{M-sp75} & 3.57 & 0.73 & 41.52 & 35.97 \\
\hspace{0.4em}\texttt{S-320}  & 4.44 & 0.82 & 44.54 & 38.27 \\
\hspace{0.4em}\texttt{S-600}  & 16.05 & 0.92 & 68.46 & 56.71 \\
\bottomrule
\end{tabular}
\end{table}

\subsection{Identity Preservation Is Domain- and Backend-Dependent}
Table~\ref{tab:asv_domains} extends ASV to all three domains with the ECAPA-TDNN backend. The architecture pattern is consistent: DAC and LanguageCodec preserve speaker identity almost everywhere, RVQ EnCodec stays low ($<3$ EER), and the low-rate tokenizers and semantic codecs (FocalCodec, WavTokenizer S-600, UniCodec, low-rate SemantiCodec) lose identity. Domain matters strongly: the strongest codecs reveal their largest EER on \textit{afrinames} (DAC rises from $0$ on the other two domains to $0.8$--$1.0$; EnCodec to $1.4$--$3.2$), confirming that the names corpus stresses speaker discrimination, whereas for the weak tokenizers all three domains are difficult and the hardest domain varies by codec (FocalCodec 12.5\,Hz peaks on multilingual, WavTokenizer S-600 on dialog).

\begin{table}[t]
\centering
\scriptsize
\setlength{\tabcolsep}{4pt}
\caption{Speaker verification EER (\%, ECAPA-TDNN) across the three domains. Lower is better. ``--'' = condition not run.}
\label{tab:asv_domains}
\begin{tabular}{@{}lccc@{}}
\toprule
\textbf{Codec / Variant} & \textbf{afrinames} & \textbf{dialog} & \textbf{multiling.} \\
\midrule
DAC \texttt{8kbps}        & 0.96 & 0.00 & 0.00 \\
DAC \texttt{16kbps}       & 0.75 & 0.00 & 0.00 \\
DAC \texttt{24kbps}       & 0.95 & 0.00 & 0.00 \\
EnCodec \texttt{3kbps}    & 3.18 & 2.04 & 2.00 \\
EnCodec \texttt{6kbps}    & 1.93 & 1.81 & 1.00 \\
EnCodec \texttt{12kbps}   & 1.39 & 0.40 & 1.00 \\
EnCodec \texttt{24kbps}   & 1.98 & 0.30 & 1.00 \\
FocalCodec \texttt{12.5Hz}& 14.67 & 10.63 & 17.78 \\
FocalCodec \texttt{25Hz}  & 8.00 & 6.42 & 8.19 \\
FocalCodec \texttt{50Hz}  & 4.81 & 4.08 & 6.00 \\
LanguageCodec \texttt{bw0}& -- & 0.00 & 0.03 \\
LanguageCodec \texttt{bw3}& -- & 0.00 & 0.04 \\
SemantiCodec \texttt{0.31kbps} & 8.67 & 3.61 & 7.00 \\
SemantiCodec \texttt{1.40kbps} & 3.19 & 0.11 & 1.00 \\
UniCodec \texttt{6.6kbps} & 4.53 & 10.20 & 2.00 \\
WavTokenizer \texttt{M-sp75}   & 4.00 & 3.57 & 2.00 \\
WavTokenizer \texttt{S-320}    & 6.00 & 4.44 & 8.00 \\
WavTokenizer \texttt{S-600}    & 12.00 & 16.05 & 13.00 \\
\bottomrule
\end{tabular}
\\[2pt]
\footnotesize Representative configurations shown; the full dialog sweep is in Table~\ref{tab:dialog_main}. LanguageCodec was not run on \textit{afrinames}; DAC on \textit{afrispeech\_multilingual} had few enrollable speakers (64 trials vs.\ $\geq$2.4k elsewhere).
\end{table}

\noindent\textbf{The ASV backend is itself an evaluation choice.} Re-scoring the same dialog conditions with a classical MFCC--GMM backend changes both the numbers and the ranking. MFCC makes WavTokenizer S-600 look benign (EER $2.4$ vs.\ $16.1$ under ECAPA) while making FocalCodec look catastrophic (EER $14$--$27$ vs.\ $4$--$11$). A weak backend therefore masks tokenizer identity loss and exaggerates FocalCodec's, flipping the apparent ranking. This extends our metric-validity point from signal metrics to the verification model: ASV conclusions are only as trustworthy as the embedding backend, and a strong neural backend is necessary.

\subsection{The Degradation Is Partly Recoverable}
\label{sec:ft}
Finally, the codec-induced ASR loss is not fixed. Adapting the codec with LoRA on African-accented speech (Sec.~\ref{sec:adapt}), while holding the recognizer fixed, lowers WER toward the uncompressed baseline on a held-out conversational subset. Table~\ref{tab:ft} reports DAC at 8\,kbps: relative to the pretrained codec, parameter-efficient adaptation reduces WER using only $1$--$3\%$ trainable parameters and is competitive with full fine-tuning. The effect is consistent across recognizers but modest in absolute terms, as DAC is already a strong RVQ codec; which architectures adapt reliably---and why convolutional decoders require care---is taken up in companion work~\cite{anon_lora}. For the benchmark, the takeaway is narrower but useful: codec choice and codec adaptation should be considered jointly, and adaptation-aware evaluation belongs alongside the signal and task metrics above.

\begin{table}[t]
\centering
\footnotesize
\caption{Codec adaptation on a held-out \textit{afrispeech\_dialog} subset (49 files), DAC 8\,kbps. WER (\%): pretrained codec vs.\ parameter-efficient (LoRA) and full fine-tuning, with the recognizer held fixed. Lower is better ($\downarrow$).}
\label{tab:ft}
\setlength{\tabcolsep}{4pt}
\begin{tabular}{@{}lcccc@{}}
\toprule
\textbf{Recognizer} & \textbf{Pretrained} & \textbf{Enc-LoRA} & \textbf{Full-LoRA} & \textbf{Full FT} \\
\midrule
Whisper-lg-v3 & 21.21 & \textbf{17.81} & 18.94 & 19.35 \\
parakeet-tdt  & 12.48 & 11.48 & \textbf{11.42} & 11.54 \\
\bottomrule
\end{tabular}
\\[2pt]
\footnotesize Trainable params: Enc-LoRA $\approx$1\%, Full-LoRA $\approx$3\%, Full FT 100\%.\\ Uncompressed Whisper $\approx$17.3 WER; parakeet $\approx$11.4 WER.
\end{table}

\subsection{Implications for Codec Selection and Inclusivity}
These results suggest a practical selection rule. For transcription on conversational African-accented speech, mid-bitrate RVQ codecs are the most stable, and very-low-bitrate semantic compression can be prohibitively destructive; if speaker identity must be preserved, ASV becomes decisive, since some ASR-usable codecs collapse verification. Rate also matters but does not tell the whole story: across dialog codecs, bitrate/token-rate is negatively correlated with WER ($\rho=-0.76$, $p=3.1{\times}10^{-5}$, $n=23$), yet architectures at similar rates differ sharply in identity and prosody preservation. Because the most predictive signal metric is task- and domain-specific, codec selection should report ViSQOL, STOI, and F0-RMSE against the target task rather than a single perceptual score. Finally, our adaptation results indicate that compression need not entrench performance gaps: lightweight tuning on representative African speech recovers most of the ASR loss, so compression can become an enabling layer rather than an additional source of disparity---provided evaluation is task-aware, domain-representative, and adaptation-aware.

\section{Limitations}
Our primary ASV backend is ECAPA-TDNN; we report MFCC--GMM only to show backend sensitivity and do not include further backends such as WavLM. ASR breadth is deepest on conversational dialog; per-domain ASR for the names and multilingual sets is limited by recognizer coverage and transcript availability, so we report those domains primarily through signal and ASV metrics. Two cells are under-powered or absent (LanguageCodec was not run on \textit{afrinames}; DAC's multilingual ASV used only 64 trials). F0-RMSE is computed on smaller matched subsets than NISQA/UTMOS/ViSQOL, so its high correlations should be interpreted as strong evidence for prosodic sensitivity rather than as a complete replacement for reference intelligibility metrics. The adaptation results summarized here cover DAC on conversational speech and are developed fully in companion work; we include them only to show recoverability, not as this paper's contribution. We do not yet include a non-African reference set, so claims about \emph{relative} disparity are deliberately conservative.

\section{Conclusion}
We benchmarked seven neural audio codecs across three African speech datasets, measuring signal quality, ASR, and ASV. Rather than assuming that perceptual scores capture utility, we measured the link and found that reference-based intelligibility/structure metrics and F0 error track downstream behavior more reliably than neural MOS predictors, with the most informative metric depending on task and domain. ASR and ASV failure modes diverge across architectures, degradation is largest for conversational speech, and parameter-efficient codec adaptation recovers part of the compression-induced ASR gap. Together these findings argue for task-aware, domain-representative, and adaptation-aware evaluation as a prerequisite for inclusive speech-compression technology.

\section*{Acknowledgment / AI-Generated Content Disclosure}
AI-assisted tools were used for limited support in code debugging, LaTeX troubleshooting, and writing clarity. All experimental design, model training, data processing, quantitative analyses, interpretation, and final claims were made and verified by the authors; no AI tool generated or altered the reported experimental results.

\end{document}